\documentclass[12pt]{iopart}
\usepackage[dvips]{graphicx}

\begin{document}
\title[The 3-SAT problem with large number of clauses
in the $\infty$-RSB scheme]
{The 3-SAT problem with large number of clauses in the 
$\infty$-replica symmetry breaking scheme}

\author{A. Crisanti$^{1}$\footnote{andrea.crisanti@phys.uniroma1.it}, 
	L. Leuzzi$^{2}$\footnote{leuzzi@science.uva.nl} and
	G. Parisi$^{1}$\footnote{giorgio.parisi@roma1.infn.it}}

\address{$^{1}$Dipartimento di Fisica, Universit\`a di Roma
	 `La Sapienza'', P.le Aldo Moro 2, I-00185 Roma, Italy \\
         and Istituto Nazionale Fisica della Materia, Unit\`a di Roma}

\address{$^{2}$ITFA and FOM,  Universiteit
         van Amsterdam Valckenierstr. 65, 1018 XE Amsterdam 
	(The Netherlands)}	

\date{\today}

\begin{abstract}
In this paper we analyze the structure of the UNSAT-phase of the 
over-constrained $3$-SAT model by studying the low temperature
phase  of the associated
disordered spin model. We derived the full Replica
Symmetry Breaking (RSB) equations  for a general class of disordered
spin models which includes the Sherrington-Kirkpatrick (SK) model,
the Ising $p$-spin model as well as the 
over-constrained $3$-SAT model as particular cases.
We have numerically solved the $\infty$-RSB
equations using
a pseudo-spectral code down to and including zero temperature.
We find that the UNSAT-phase of the over-constrained $3$-SAT
is of the $\infty$-RSB kind: in order to get a stable solution
the replica symmetry has to be broken in a continuous way, similarly 
to the SK model in external magnetic field.

\end{abstract}

\maketitle

\section{Introduction}
A  combinatorial optimization problem 
is defined, in a  broad sense,
 by specifying
a certain number of
free variables constituting it
and the conditions that  its solution  must satisfy.
In treating an optimization problem the fundamental step is to find the
{\em most 
efficient} algorithm yielding the solution.
Efficient from the point of view of all the computing resources needed
for its performance,
the most important of which is the time requirement, and, in particular,
its dependence on the {\em size} of the problem. By size we mean, in an
informal way, the number of variables, or, even,
 the number of conditions.

A computational hard problem \cite{garey,papa}
is an optimization problem 
for which the time needed to find the solution, or 
to determine with  certainty that it has no solution,
 very sensitively increases with the size. 
More specifically there are no 
polynomial algorithms able to solve it.
This kind of problems are therefore called {\em intractable} or
Non Polynomial (NP).

Any NP problem can be reduced to a particular NP paradigmatic
problem 
(the SAT problem) exploiting algorithms performing such a  mapping
 in a polynomial time 
\cite{Cook71,karp}.
Computer  scientists call  this class of very hard combinatorial 
optimization problems
the class of  Non Polynomial-Complete (NPC) problems \cite{garey}.

Summing-up,
a  NPC problem is defined, in a qualitative way, as an optimization
 problem
whose solutions, or the certainty that it has no solution, can only be 
found, in the worst case, by algorithms whose computation time
grows faster than any polynomial with the number of variables of the system.

In this paper we shall consider the 3-SAT problem \cite{karp}. 
This is a particular 
version of SAT problem, which is the paradigm of NPC combinatorial problems 
showing a phase transition.
For this problem the free variables are boolean variables and the conditions 
are sets of three boolean numbers.
Its importance, apart from being a historical one,
comes from the fact that even if theoretically
any NPC problem can be mapped in any other, in practice given problems are 
better suited for proving such correspondence. 
Of these reference problems one of the most 
useful (and used) is the 3-SAT.

In the last years a one to one correspondence has been observed
between computational hard problems
 and the ground state properties of spin-glass 
models \cite{MZPRL96,MZPRE97}. 
Statistical mechanics has been 
 applied to the study of  universal behaviour 
in the computational cost of some class of algorithms,
searching for solutions  of random realizations of the prototype
of the NPC problems: the Satisfaction (SAT) problem \cite{SK96}.
The investigation of the properties of NPC problems is then performed 
through the introduction of an energy or cost function and an 
artificial temperature. In this mapping the actual NPC problem
is recovered as the $T=0$ limit of the associated statistical mechanical 
problem.
Such an approach has  been implemented both numerically, using
simulated annealing algorithms \cite{KGV83}), 
and analytically.

To set up a statistical mechanical approach
one first introduces a semidefinite positive  Hamiltonian ${\cal{H}}[C]$
function, defined for each
given instance $C$ of the problem, 
constructed in such a way that if the configuration $C^{*}$ 
is solution of the computational problem then
${\cal{H}}[C^{*}]=0$. On the contrary, 
if ${\cal{H}}[C]>0$ for any $C$ then problem 
does not admit solution.  Having defined an Hamiltonian, the associated
statistical mechanical problem is described by the partition function
\begin{equation}
Z(\beta)=\sum_{C}\exp(-\beta \cal{H}[C]) \ ,
\end{equation}
where $\beta^{-1} =T$ is the (artificial) temperature of the system.
Proceeding further one introduces the 
usual thermodynamic quantities, e.g., the energy
\begin{equation}
U(\beta)=-{ \partial \ln(Z(\beta))\over \partial \beta} \ .
\end{equation}
The mapping is not trivial since intensive quantities, such as
the energy density $u\equiv U/N$, do not depend 
on $N$ in the infinite $N$ limit, so that a computation of 
their average over the distribution of instances 
of the computational problem is sufficient to obtain 
relevant informations on its satisfiability.

Eventually, to recover the original computational problem, 
the limit $T\to 0$ has to be taken. We stress that, in this approach,
the temperature only plays a role for  the constsruction
of a statistical mechanics problem, of which the only interesting features are
those at $T=0$. 

In this approach, phase transition concepts play
an important role to  build a theory for typical case complexity in 
theoretical computational science.
The importance comes from the fact that NPC decision problems, that are 
computationally hard in the worst case,
may be not in the typical case, if one excludes the critical regions of the 
parameter space, where almost all instances become 
computationally hard to solve. 
The computational critical region corresponds to a phase transition region
in statistical mechanics  language.
Far from phase boundaries the problems are either under-constrained
or over-constrained and  one can determine search procedures 
able to  find solutions, or certainty of no solution, in polynomial times:
 the results of worst-case complexity theory are not very relevant
 in practice and what is necessary is
 a theory for typical-case  complexity.
To such a purpose, the analysis of general search 
methods applied to different classes of hard computational problems, 
characterized by a large number or relevant, randomly generated, variables,
is fundamental.

Variables are under-constrained when the minimal number of
violated clauses does not depend on their possible assignments.
In particular this is true when they do not appear in any clause.
In the under-constrained phase the clauses of the problem can always be
satisfied (SAT phase).
On the contrary, 
variables are over-constrained when they cannot satisfy simultaneously 
all the clauses imposed on them. In this case we are in the UNSAT phase.
Going back to the mapping onto a statistical mechanical problem, 
the UNSAT-phase corresponds to a frozen (spin-glass) 
phase while  the SAT-phase corresponds 
to an  ordered (ferromagnetic) phase.

The 3-SAT problem, and, in general, the $K$-SAT problem where the 
clauses contain a number $K$ of elements,
 can be mapped onto a diluted long-range 
spin-glass model \cite{MZPRL96,MZPRE97}.
The model is mean-field 
because of the lack of geometrical correlations in the clauses.
However, since each spin has only a finite number of neighbours 
strong local field fluctuations, stronger than in those of
fully connected spin-glass, are present.

The relevant parameter driving the SAT/UNSAT transition is the ratio 
$\alpha$
between clauses and number of variables of the system, 
which is the  connectivity in the statistical mechanical analogue 
of the combinatorial problem.

Indeed too many 
conditions cause the  unsatisfiability of the problem.
The entropy of the associated spin model gives  a measure of the typical 
number of solutions. Therefore at the transition an abrupt disappearance 
of all (exponentially numerous) solutions makes the entropy jump to zero. 

For  $K=1$ and $2$ the problem is solvable: the time to find
the solutions grows polynomially (actually even linearly \cite{Cook71})
with the number of variables.
For $K\geq 3$ the problem is, on the contrary, NPC.
The transition threshold for 3-SAT  has been determined numerically
at $\alpha_c(3)\simeq 4.25 - 4.30$ \cite{KSscience}.

Based on the mapping of random clauses onto the quenched disorder
of the associated spin model, 
in Ref. \cite{MZPRL96,MZPRE97} the Replica trick was introduced
to compute the statistical mechanics of the K-SAT problem 
and the Replica Symmetric (RS) theory was carried out.
The K-SAT problem is naturally mapped onto a disordered spin model
with finite connectivity, where
the role of connectivity is played by the density of clauses.
Even if it gives a qualitative good pattern of the transition it is, 
however, unable to predict correctly the value of the transition 
threshold between the SAT-phase and the UNSAT-phase and
the correct thermodynamic quantities in the UNSAT-phase.

The failure of the RS solution can be traced back to the
existence of a very large number of equilibrium states of the
associated statistical mechanical problem in the thermodynamic limit 
$N\to\infty$.
To deal with those, and improve the knowledge
of the structure of the solutions of the decision problem,
 it is necessary to break the Replica Symmetry.
Replica Symmetry Breaking  (RSB)  in diluted models is
a very hard issue, due to the complex
 structure of the saddle point equations, 
for recent approaches see \cite{MP1RSB,FRTZ01}.

As shown in \cite{nature} the SAT/UNSAT transition results from  the
sudden freezing of a finite number of variables, as $\alpha$ increases
above $\alpha_c$. These variables form a {\em backbone}
that does not disappear in the thermodynamic limit.
Information about the structure of this backbone and about 
the mutual overlap between different assignments minimizing the cost function
in the UNSAT phase is, then, very important to understand the transition.

In an attempt to overcome the difficulty of solving a spin glass
diluted model a
variational approach, both for RS and $1$-RSB solution, 
has been recently proposed.  It is based on the existence of the backbone of
over-constrained variable that remains finite in the thermodynamic limit
\cite{BMW}. This circumvent the necessity of solving
the self-consistent equations for RS and $1$-RSB, but does not resolve
the question about the nature of the RSB solution in the
UNSAT-phase. Thus  the relation between RSB transition and 
the typical case complexity theory is yet an open question.

In order to investigate the the nature of the RSB solution in the
UNSAT-phase, in this paper we shall consider the K-SAT problem with $K=3$, 
the simplest NPC problem of this class, in the limit of a large number 
of clauses $\alpha\gg 1$ (over-constrained), 
where the associated statistical mechanical 
problem can be handled with known techniques of disordered spin
systems.
The basic idea is that since there should be no
other transitions for $\alpha > \alpha_c$, the structure of
the UNSAT-phase for $\alpha\gg \alpha_c$ should be representative of
the whole UNSAT-phase in the range $\alpha > \alpha_c$.
Performing a careful study of the thermodynamic quantities
down to zero temperature of the associated disordered spin model
we find that the RSB is infinitely broken ($\infty$-RSB)
in the UNSAT-phase.

Moreover, using the first
two terms in the asymptotic expansion in $1/\sqrt{\alpha}$
of the thermodynamic quantities, 
we can obtain an upper bound for $\alpha_c$.

The paper is organized as follows.
In Section \ref{secMod} we introduce the model.
The details of the derivation of the $\infty$-RSB solution 
with an arbitrary gauge
are  given in section {\ref{secIRSB}}. The calculation is carried out
for a general model introduced in 
Section \ref{secR}, to which the over-constrained $3$-SAT problem belongs.
Here we also sketch the procedure used for the numerical
of solution of the $\infty$-RSB solution.
The $\infty$-RSB solution for the over-constrained $3$-SAT model
is discussed in Sections \ref{IRSB} and \ref{TD}.

\section{The over-constrained 3-SAT model}
\label{secMod}

The model we study has been introduced in  \cite{LPJSP01},
where the analysis of RSB at one and two steps has been carried out.
The $3$-SAT model is defined by a set of boolean variables  $s(i)=0,1$, 
defined on the sites $i=1,\ldots, N$, and an ensemble of randomly 
generated $3$-SAT boolean formulae. 
First the random boolean formulae are constructed by 
assign to each triplet $\{i_1,i_2,i_3\}$, with $i_1<i_2<i_3$, 
a set of three  independent variables 
$\epsilon_{1,2,3}$ which take the 
value $+1$ or $-1$ with probability $1/2$.
Next for each instance of the problem triplets of randomly chosen sites 
$\{i_1,i_2,i_3\}$ are selected by assigning to the 
variables $r_{i_1,i_2,i_3}$ the value $1$ with probability 
$p\equiv \alpha N^{-2}$ and $0$ with probability $1-p$.
For $N\to \infty$ there are $\alpha N$ variables $r$ which are 
different from zero, and hence $\alpha N$ $3$-SAT boolean formulae..

If we introduce the spin variables $\sigma(i)=1-2 s(i)$, the cost 
function reads
\begin{equation}
\fl {\cal{H}}=\sum_{i_1<i_2<i_3}r_{i_1,i_2,i_3}
\frac{1-\epsilon_{1}^{(i_1,i_2,i_3)}\sigma(i_1)}{2}
\frac{1-\epsilon_{2}^{(i_1,i_2,i_3)}\sigma(i_2)}{2}
\frac{1-\epsilon_{3}^{(i_i,i_2,i_3)}\sigma(i_3)}{2}.
\label{HAMILTONIAN}
\end{equation}
which is nothing but the number of unsatisfied clauses. Indeed 
it is easy to see that  each term is either $1$ (unsatisfied)
or $0$ (satisfied). Note that  
${\cal{H}}=0$ if and only if all the clauses are satisfied.

The statistical mechanical approach to the $3$-SAT problem 
takes ${\cal{H}}$ as the Hamiltonian of a disordered spin system,
and, as discussed above, studies the properties of the ground state.

The fundamental quantities in studying the hard optimization 
problems with the tools of statistical mechanics are the zero temperature 
energy and entropy densities, respectively $u_0$ and $s_0$, in the 
thermodynamic limit.
$u_0$ represents the average over the distribution of clauses of
the number of clauses that are not satisfied by the formula 
(\ref{HAMILTONIAN}). $s_0$ is the logarithm of the number of
solutions satisfying the formula divided by the number of variables.
For the behaviour of these quantities 
as functions of the connectivity-like parameter  $\alpha$,
the conjecture is done\cite{MZPRL96,MZPRE97} that
\begin{eqnarray}
u_0(\alpha)=0 \ , && \hspace*{1 cm} s_0(\alpha)>0, \hspace*{2 cm}
\mbox{for $\alpha<\alpha_c$}
\\
u_0(\alpha)>0 \ , && \hspace*{1 cm} s_0(\alpha)=0, \hspace*{2 cm}
\mbox{for $\alpha>\alpha_c$}
\end{eqnarray}

For $\alpha \ll \alpha_c$ the problem is quite 
under-constrained and it is relatively easy to find an assignment 
of variables $\sigma_{i}$ 
satisfying the clauses.  In other words for $\alpha < \alpha_c$
the problem is SAT, 
with probability going to 1 for $N\to\infty$, 
On the contrary for $\alpha > \alpha_c$ the problem
does not have solutions, UNSAT-phase.
The analysis of the UNSAT-phase is in general rather hard.
The most difficult case occurs around
$\alpha_c$ where an exponential time may be needed to determine
the unsatisfiability.
Away from the critical region, i.e.,  $\alpha \gg \alpha_c$, to prove 
unsatisfiability is easier, and more insight into the
structure of the phase space can be gained.

In the present paper  we will work in the over-constrained approximation
$\alpha \gg\alpha_c$, where the computation strongly simplifies. 
This limit is obtained by expanding in $1/\sqrt{\alpha}$ to second order,
after having rescaled the temperature
$\beta \to \beta / \sqrt{\alpha}$ .
For further details See \cite{LPJSP01}.

Note that in Ref. \cite{LPJSP01} the reduced inverse temperature
was $\beta=\mu/\sqrt{\alpha}$.  Here we will, instead,
keep the notation $\beta$ also for the reduced inverse temperature.
Moreover in \cite{LPJSP01} the clauses are erroneously over counted in 
the evaluation of the partition function \cite{Zul}.
This, however, 
does not produce any relevant change, apart from a rescaling of the reduced 
temperature
and of the energy and the free energy
of a factor $1/\sqrt{6}$,  leaving the entropy invariant.{\footnote[1]{In  
\cite{LPJSP01} the reduced inverse temperature
was called $\mu$, therefore the substitution $\mu=\sqrt{6}\,\beta$ 
cure the difference.}}
In order to make a comparison with the results shown there,
it is enough to multiply $\beta$, the free energy and the energy shown 
in the present  paper
by a factor $\sqrt{6}$.


\section{The replica approach in a generalized form}
\label{secR}
The 3-SAT model belongs to the family of spin models interacting via
quenched random couplings. 
These are described by a random Hamiltonian 
${\cal H}[J;\sigma]$ where $J$ are the random ``quenched'' couplings.
For example, in the Sherrington-Kirkpatrick (SK) model $J$ is a 
symmetric Gaussian
matrix of zero mean and variance proportional to $1/N$ \cite{SKPRL75}, while
in its $p$-spin generalization the variance goes like $1/N^{p-1}$ \cite{Gardner}.
For the 3-SAT problem the disorder is introduced by the random clauses 
imposed 
on the set of variables. In the simple limit that we are considering here, 
the quenched disorder is represented by the random $\pm 1$ variables
$\epsilon^{(i_1,i_2,i_3)}$, assigning a clause on the three
sites $i_1$, $i_2$ and $i_3$ \cite{LPJSP01}.

For any fixed coupling realization $J$, the partition function of the
spin system, with $N$ spins, is given by \cite{MPV,FH}
\begin{equation}
   Z_N[J]= \mbox{Tr}_{\sigma}\ \exp\bigl(-\beta {\cal H}[J;\sigma]\bigr)
\end{equation}
and the quenched free energy per spin is
\begin{equation}
f_N= -{1\over N\beta}\ \overline{\,\ln Z_N\,}
        = -{1\over N\beta}\ \int d [J]\ P[J]\ \ln Z_N[J]
\end{equation}
where $\overline{(\cdots)}$ indicates the average over the couplings
realizations. We assume that the thermodynamic limit of the free energy,
$-\lim_{N\to\infty} \ln Z_N[J]\ /\ N\beta$
is well defined and is equal to the quenched free energy
$f=\lim_{N \to \infty} f_N$
for almost all coupling realizations $J$ (self-average property).

The analytic computation of the quenched free energy, i.e., of the average of
the logarithm of the partition function, is a quite difficult problem,
even in simple cases as nearest neighbour one dimensional models.
However, since the integer moments of the partition function are easier
to compute, the standard method uses the so called ``replica trick''
by considering the annealed free energy $f(n)$ of $n$ non-interacting
`replicas' of the system \cite{SKPRL75,MPV,FH},
\begin{equation}
f(n)= -\lim_{N\to\infty} {1\over N\beta n}\ \ln\left[
                                          \overline{(Z_N[J])^n}
                                               \right].
\end{equation}
The quenched free energy of the original system is then recovered as the
continuation of $f(n)$ down to the unphysical limit $n=0$,
\begin{equation}
   f= -\lim_{N\to \infty}\lim_{n\to 0}
                         {\overline{(Z_N[J])^n} - 1\over N\beta n}
      = \lim_{n\to 0} f(n).
\end{equation}
In the last equality we assumed that the replica limit and the
thermodynamic limit can be exchanged.
This procedure replaces the original interactions in the real space
with couplings among different replicas. 
The interested reader can find a complete and detailed 
presentation of the replica method for disordered statistical mechanical 
systems in Ref. \cite{FH} and in Ref. \cite{MPV}.

In what follows we shall consider disordered spin systems for which 
$f$ in the replica space can be written in the form
\begin{eqnarray}
\fl \beta f\left[Q_{ab},\Lambda_{ab}\right] =
	 \beta f_0(\beta)
	-\frac{\beta^2}{2}\lim_{n\to 0}
             \frac{1}{n}\sum_{a\neq b}^{1,n}g\left(Q_{ab}\right)
	+\frac{\beta^2}{2}\lim_{n\to 0}
       \frac{1}{n}\sum_{a\neq b}^{1,n}\Lambda_{ab}Q_{ab}
\nonumber \\
\lo-\lim_{n\to 0}\frac{1}{n}\log\mbox{Tr}_{\sigma}
\exp\left(\frac{\beta^2}{2}\sum_{a\neq b}\Lambda_{ab}\,\sigma^a\sigma^b\right)
\label{fen_Qab}
\end{eqnarray}
where $Q_{ab}$ is the spin-overlap matrix in the replica space 
between replicas
$a$ and $b$:
\begin{equation}
	Q_{ab} = \frac{1}{N}\sum_{i=1}^{N}\, 
	\overline{ \langle\sigma_i^a\,\sigma_i^b\rangle}
\label{Qab}
\end{equation}
and $\Lambda_{ab}$, the Lagrange multiplier associated with $Q_{ab}$,
gives the interaction matrix between spins of different replicas.
Angular brackets denote thermal average.
Stationarity of $f$ with respect to variations of $\Lambda_{ab}$ and
$Q_{ab}$ leads to the  self-consistency equations for 
the matrices $\Lambda$ and $Q$:
\begin{eqnarray}
\Lambda_{ab} &=& g_1\left(Q_{ab}\right)
\\
Q_{ab}&=&\frac{ \mbox{Tr}_{\sigma}\ \sigma^a\sigma^b
	        \exp\left(\frac{\beta^2}{2}
                  \sum_{a\neq b}\Lambda_{ab}\sigma^a\sigma^b\right)
              }
              {\mbox{Tr}_{\sigma}\exp\left(\frac{\beta^2}{2}
                \sum_{a\neq b}\Lambda_{ab}\sigma^a\sigma^b\right)}
\end{eqnarray}
where we have used the short-hand notation
\begin{equation}
g_n(z)\equiv\frac{d^ng(z)}{d z^n} \qquad  n=1,2,\ldots
\end{equation}

The function $g$ and the constant $f_0$  depend on the specific model.
For example for the SK model
we have\cite{SKPRL75}:
\begin{equation}
 g(z)=\frac{z^2}{2}, \hspace*{1 cm} f_0=-\frac{\beta}{4}.
\label{defgSK}
\end{equation}
Similarly the $p$-spin model \cite{Gardner} is recovered for
\begin{equation}
 g(z)=\frac{z^p}{2}, \hspace*{1 cm} f_0=-\frac{\beta}{4}
  \hspace*{0.5 cm}
\label{defgp}
\end{equation} 
Finally, for the $3$-SAT problem in the limit
of low dilution  we have
{\footnote[2]{In 
\cite{LPJSP01} it
 was $f_0=-\frac{\mu}{16}+\frac{\sqrt{\alpha}}{8}$ and 
$g(z)=\frac{1}{64}(1+z)^3$. Due to the over-counting discussed 
at the end of the introduction (i.e. $\mu f=6 \beta f$),
 these  were six times bigger than the
 actual definition.}}
\cite{LPJSP01}:
\begin{equation}
 g(z)=\frac{1}{384}(1+z)^3, \hspace*{0.5 cm} f_0=-\frac{\beta}{96}
      +\frac{\sqrt{\alpha}}{48}
\label{defg3}
\end{equation}
In the following we will use $f_0=-\beta/96$,  the only consequence being 
a shifting of the free energy density $f$
and of the internal energy density $u$ by a factor $\sqrt{\alpha}/48$.


\section{Infinite Replica Symmetry Breaking Solution}
\label{secIRSB}

\subsection{$\infty$-RSB solution}
To evaluate the $n\to 0$ limits in (\ref{fen_Qab}) one has to make an
{\it Ansatz} on the structure of matrices $\Lambda$ and $Q$, i.e., to choose
a Replica Symmetry Breaking (RSB) scheme.
In order to be as general as possible, we shall use the RSB scheme introduced
by de Dominicis, Gabay and Orland \cite{DGOJPL81,DGDJPA82}, which besides
the Edwards-Anderson order parameter \cite{EA} also involves the
anomaly to the linear response function, also called Sompolinsky's anomalies
\cite{SPRL81}. The more usual Parisi's RSB scheme is recovered by 
a proper gauge fixing. 
Here we shall only report the main results, since the calculation is 
straightforward. The interested reader can find some details 
in Ref. \cite{DGOJPL81,DGDJPA82}. 

By applying the RSB scheme infinite times and introducing two functions
$q(x)$ and $\lambda(x)$, $0\leq x\leq 1$, one for each matrix, 
 the free energy functional 
(\ref{fen_Qab}) becomes \cite{DGOJPL81,DGDJPA82}:
\begin{eqnarray}
\fl\beta f(\beta)= \beta f_0(\beta)+\frac{\beta^2}{2}\left[g\left(q(1)\right)+
\lambda(1)\left(1-q(1)\right)\right]
\label{f1}
+\frac{\beta}{2}\int_0^1 dx \ g_1\left(q(x)\right) \dot\Delta_q(x)
 \\ \nonumber
\hspace*{- 1 cm}-\frac{\beta}{2}\int_0^1 dx \left[q(x) \dot\Delta_{\lambda}(x)
+\lambda(x) \dot\Delta_q(x)\right]
-\beta\int_{-\infty}^{+\infty} \frac{d y}{\sqrt{2 \pi \lambda(0)}}
\exp\left(-\frac{y^2}{2\lambda(0)}\right)\phi(0,y)
\end{eqnarray}
where $\phi(0,y)$ is the solution evaluated at $x=0$ 
of the the Parisi's equation
\begin{equation}
\dot\phi(x,y)=-\frac{\dot{\lambda}(x)}{2}\,\phi''(x,y)
+\frac{\dot\Delta_{\lambda}(x)}{2}\,\phi'(x,y)^2 
\label{eqPhi}
\end{equation}
with the boundary condition
\begin{equation}
\phi(1,y)=T\log\left(2\cosh \beta y\right)
\label{Phi1}
\end{equation}
and $\Delta_q(x)$ and $\Delta_{\lambda}(x)$ are the anomalies 
associated with the order parameters $q(x)$ and $\lambda(x)$.
We have used the standard notation and denote derivatives with respect to 
$x$ by a dot and derivatives with respect to $y$ by a prime. Note that
with this notation Sompolinsky's $\Delta'$ becomes $T\dot\Delta$.
It is easy to see that using (\ref{defgSK}) one recovers the Sompolinsky
functional for the SK model \cite{SPRL81}, and inserting the
Parisi's gauge $\dot\Delta_q(x) = -\beta x\dot q(x)$ the 
Parisi's functional \cite{Pphieq}.

The Parisi's equation (\ref{eqPhi}) can be included into the free energy via the 
Lagrange multiplier $P(x,y)$ and the initial condition at $x=1$ (\ref{Phi1})
via $P(1,y)$. The free energy then becomes \cite{SDJPC84}
\begin{eqnarray}
\fl \beta f_v(\beta)= \beta f(\beta)
  +\beta \int_{-\infty}^{+\infty} dy\  P(1,y)\,
          \bigl[\phi(1,y)-T \log\left(2\cosh \beta y\right)\bigr]
\label{eqfrev}
\\ \nonumber
\lo-\beta\int_0^1dx\int_{-\infty}^{+\infty} dy\ P(x,y) \left[\dot\phi(x,y)+
\frac{\dot{\lambda}(x)}{2}\phi''(x,y)
-\frac{\dot\Delta_{\lambda}(x)}{2}\left(\phi'(x,y)\right)^2 
\right].
\end{eqnarray}
By this construction $f_v$ is stationary with respect to  variations of
$P(x,y)$, $P(1,y)$, $\phi(x,y)$, $\phi(0,y)$, the order parameters
$q(x)$ and $\lambda(x)$ and anomalies 
$\dot\Delta_q(x)$ and $\dot\Delta_\lambda(x)$.
Variations with respect to $P(x,y)$ and $P(1,y)$ simply give back 
eqs. (\ref{eqPhi}) and (\ref{Phi1}).
Stationarity with respect to variations of $\phi(x,y)$ and $\phi(0,y)$
leads to a partial differential equation for $P(x,y)$:

\begin{equation}
\dot P(x,y) =\frac{\dot\lambda(x)}{2}\, P''(x,y)+
     \dot\Delta_{\lambda}(x)\, \bigl[P(x,y)\,\phi'(x,y)\bigr]'.
\label{eqP}
\end{equation}
with the boundary condition at $x=0$
\begin{equation}
P(0,y)=\int_{-\infty}^{+\infty}\frac{d y}{\sqrt{2 \pi \lambda(0)}}
\exp\left(-\frac{y^2}{2\lambda(0)}\right).
\label{P0}
\end{equation}
Finally, variations of  $q(x)$, $\dot\Delta_q(x)$, $\lambda(x)$
and $\dot\Delta_{\lambda}(x)$ lead to
\begin{equation}
\dot\Delta_{\lambda}(x)=g_2\left(q(x)\right)\dot\Delta_q(x)
\label{Dla}
\end{equation}
\begin{equation}
 \lambda(x)=g_1\left(q(x)\right)
\label{la}
\end{equation}
\begin{equation}
\Delta_q(x)=-\beta\bigl[1-q(1)\bigr]+
\int_{-\infty}^{\infty} dy\, P(x,y)\, \phi''(x,y)
\label{Dq}
\end{equation}
\begin{equation}
q(x)=\int_{-\infty}^{\infty} dy\, P(x,y)\, \phi'(x,y)^2
\label{q}
\end{equation}
with  $\Delta_q(1)=0$, the anomalies at the shortest time-scale, corresponding to $x=1$, being zero by construction.

The Lagrange multiplier $P(x,y)$ gives the probability distribution
of local fields. One may indeed 
associate a given overlap $q(x)$ with a time scale 
$\tau_x$ such that for times of order  $\tau_x$ states with an overlap 
equal to $q(x)$ or greater can be reached by the system.
In this picture the $P(x,y)$ becomes the probability distribution
of frozen local fields $y$ at the time scale labeled by $x$ \cite{SDJPC84}.

By partial derivate the above expressions we can obtain some
useful relations. For example, deriving with respect to $x$ 
equations (\ref{q}) and (\ref{la}), or equivalently (\ref{Dq}) 
and (\ref{Dla}), one gets
\begin{equation}
g_2\left(q(x)\right)\int_{-\infty}^{\infty}dy\ P(x,y) \ \phi''(x,y)^2=1.
\label{rel0}
\end{equation}
A further derivation with respect to $x$ leads to 
\begin{equation}
\fl g_3\bigl(q(x)\bigr)\, \dot q(x)+g_2\bigl(q(x)\bigr)
\int^{\infty}_{-\infty} dy\, P(x,y)\, \bigl[\dot\lambda(x)\,\phi'''(x,y)^2
         +2\dot\Delta_{\lambda}(x)\,\phi''(x,y)^3\bigr]=0.
\end{equation}
Using equations (\ref{la}) and (\ref{Dla}) this becomes
\begin{equation}
\lo-\frac{\dot q(x)}{\dot\Delta_q(x)}=
-\frac{\dot \lambda(x)}{\dot\Delta_{\lambda}(x)}=
\frac{2\int dy^{\infty}_{-\infty} P(x,y)\ \phi''(x,y)^3}
{g_3\left(q(x)\right)+g_2\left(q(x)\right)
\int^{\infty}_{-\infty} dy P(x,y)\ \phi'''(x,y)^2},
\end{equation}
which determines the gauge relation between $q(x)$ and $\Delta_q(x)$ and
between $\lambda(x)$ and $\Delta_\lambda(x)$, the Parisi's $\beta x$.

Finally we note that from eq. (\ref{q}) and the interpretation of $P(x,y)$ 
distribution of local fields, $m(x,y) = \phi'(x,y)$ can be interpreted 
as the local magnetization over the time-scale $x$. It obeys the equation
\begin{equation}
 \dot m(x,y)=-\frac{\dot\lambda(x)}{2}\, m''(x,y)
 +\dot\Delta_{\lambda}(x)\, m(x,y)\ m'(x,y) \ ,
\label{eqm}
\end{equation}
with initial condition
\begin{equation}
m(1,y)=\tanh(\beta y).
\label{eqm1}
\end{equation}

In the next Section we shall report the results of numerical integration
of the above equations for the specific case of $3$-SAT in the 
limit of a large number of clauses.

\subsection{Thermodynamic quantities}

Since the free energy density $f_v$ [eq. (\ref{eqfrev})] is stationary we can 
easily calculate thermodynamic derivatives to compute for example the
the energy density $u$:
\begin{eqnarray}
\fl u =\frac{\partial}{\partial \beta}\, \beta f_v
    = \frac{\partial}{\partial \beta}\, \beta f_0
 +\beta\bigl[g\left(q(1)\right)+\lambda(1)\left(1-q(1)\right)\bigr]
 -\frac{1}{2}\int_0^1 dx \ q(x)\ \dot\Delta_{\lambda}(x)
\label{u_g}
\\ \nonumber
\hspace*{-1.5 cm}+\int_{-\infty}^{\infty} dy \ P(1,y)\ \phi(1,y)
-\int_{-\infty}^{\infty} dy \ P(0,y)\ \phi(0,y)
-\int_{-\infty}^{\infty} dy \ P(1,y)\ y \ \tanh(\beta y).
\end{eqnarray}
This expression can be simplified using the relation
\begin{equation}
\fl\int_{-\infty}^{\infty}dy\ P(1,y)\  \phi(1,y)
-\int_{-\infty}^{\infty}dy\ P(0,y)\ \phi(0,y)=
-\frac{1}{2}\int_0^1 dx\ q(x)\, \dot\Delta_{\lambda}(x)\ .
\label{rel1}
\end{equation}
which follows computing 
$\int_0^1 dx \int_{-\infty}^{\infty} dy P(x,y)\,\dot\phi(x,y)$  
using either (\ref{eqPhi}) or (\ref{eqP}) 
and  equating the results.

We can equivalently compute $u$ by taking the derivative
of the free energy density (\ref{fen_Qab}) as a function of the generic matrix
$Q_{ab}$, before any RSB scheme is introduced:
\begin{equation}
\frac{\partial}{\partial\beta}\, \beta f\left[Q_{ab},\Lambda_{ab}\right]=
f_0(\beta)+\beta\frac{\partial}{\partial \beta} f_0(\beta)
-\beta \lim_{n\to 0}\frac{1}{n}\sum_{a\neq b}^{1,n}g\left(Q_{ab}\right)
\end{equation}
By inserting now the chosen RSB scheme, and taking the $n\to 0$ limit,
we obtain the alternative form
\begin{equation}
u=f_0(\beta)+\beta\frac{\partial} {\partial\beta}  f_0(\beta)
 +\beta g\left(q(1)\right)
+\int_0^1 dx g_1\left(q(x)\right) \dot\Delta_q(x).
\label{u2_g}
\end{equation}

Note that by equating (\ref{u_g}) and (\ref{u2_g})
we get another integral relation:
\begin{equation}
\fl \int_0^1 dx \bigl[q(x)\ \dot\Delta_{\lambda}(x)
+\lambda(x)\dot\Delta_q(x)\bigr]
=-\int_{-\infty}^{\infty} dy\ P(1,y)\ y\ \tanh( \beta y)
+\beta\,\bigl[1-q(1)\bigr].
\label{rel2}
\end{equation}

Similarly we easily obtain the entropy density,
\begin{eqnarray}
 s&=&\beta^2\frac{\partial f_0}{\partial \beta}
+\frac{\beta^2}{2}\bigl[g\left(q(1)\right)+\lambda(1)\left(1-q(1)\right)\bigr]
\label{s}\\
\nonumber
&+&\beta \int_{-\infty}^{\infty} dy \ P(1,y)
\bigl[\log 2 \cosh \beta y - y \ \tanh(\beta y)\bigr].
\end{eqnarray}

\subsection{Numerical Integration of the $\infty$-RSB equations}
In order to study the low temperature regime of the $3$-SAT in the 
limit of a large number of clauses we have numerically integrated the 
$\infty$-RSB equations to determine $q(x)$, $P(x,y)$ and $m(x,y)$.
We followed the iterative scheme of Ref. \cite{SDJPC84,topomoto}, 
but with an improved numerical method which allows for very
accurate results for all temperatures. 

We start from an initial guess for $q(x)$, then $m(x,y)$, $P(x,y)$ and the
associated $q(x)$ are computed in the order as:
\begin{enumerate}
\item 
	Compute $m(x,y)$ integrating from $x=1$ to $x=0$ eqs. 
	(\ref{eqm}) with initial condition (\ref{eqm1}).
\item
	Compute $P(x,y)$ integrating from $x=0$ to $x=1$ eqs. 
	(\ref{eqP}) with initial condition (\ref{P0}).
\item
	Compute $q(x)$ using eq. (\ref{q}).
\end{enumerate}
The steps $1.\,\to\, 2.\,\to\, 3.$ are repeated until a reasonable convergence
is reached, typically mean square error on $q$, $P$ and $m$ is of the 
order $O(10^{-6})$.

The core of the integration scheme is the integration of the partial 
differential equations (\ref{eqm}) and (\ref{eqP}). 
In previous works this was carried out through 
direct integration in the real space which requires a large 
grid mesh to obtain precise results. To overcome such problems
we use a pseudo-spectral\cite{Orsz} dealiased \cite{deal} code
on a grid mesh of $N_x\,\times\, N_y$ points,
which covers the $x$-interval $[0,1]$ and the $y$-interval 
$[-y_{\rm max},y_{\rm max}]$. De-aliasing has been obtained by 
a $N/2$ truncation, which ensure better isotropy of numerical treatment.
The $x$ integration has been performed using an third-order Adam-Bashfort 
scheme. Typical values used are $N_x=100 \div 5000$,
$N_y = 512 \div 4096$ and $y_{\rm max} = 12 \div 48$. The number of iterations
necessary to reach a mean square error on $q$, $P$ and $m$ 
of  order $O(10^{-6})$ is few hundreds.  
More details can be found in Ref. \cite{Tommy}.

\section{$\infty$-RSB solution of the Highly Constrained 3-SAT problem}
\label{IRSB}

In the numerical solution of the $\infty$-RSB equations we used different
gauges depending on the temperature range. The reasons is that
the Parisi's gauge $\dot\Delta_q=-\beta x \dot q$
($\dot\Delta_{\lambda}=-\beta x \dot \lambda$), which
uses a simple relation between order parameters and anomalies,
leads to numerical
instabilities for large $\beta$ since it is coupled with a (numerical)
derivative. On the contrary, since in this gauge the derivatives goes to
zero as $x\to 1$, it is rather useful for not too large values of 
$\beta$. Typically for $T$ larger than $0.02\div 0.04$.

\begin{figure}[hbt]
\begin{center}
\includegraphics*[width=0.7\textwidth,height=0.5\textwidth]
{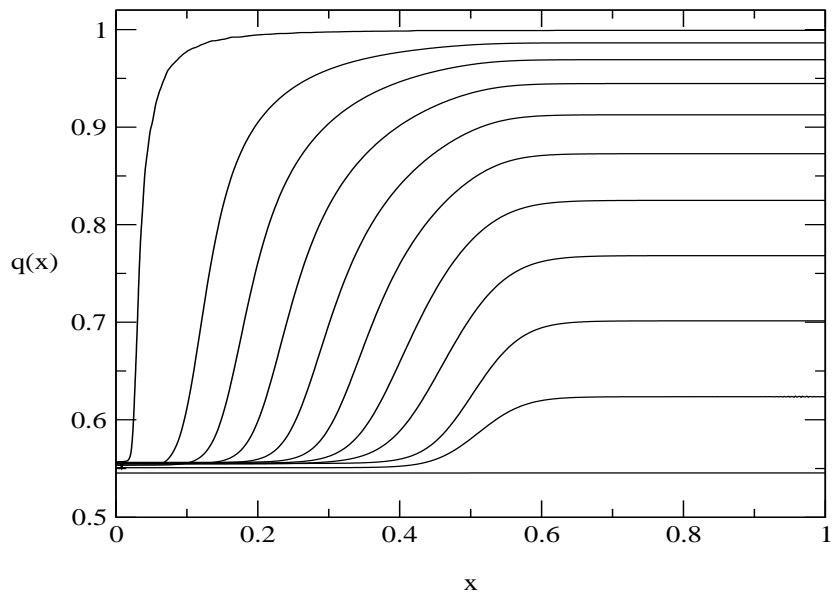}
\end{center}
\caption{$q(x)$ for the 3-SAT model 
	 for temperatures (top to bottom)
	 $T= 0.0041$, $0.0163$, $0.0245$, $0.0327$, $0.0408$, 
	$0.0490$, $0.0572$, $0.0653$,	
         $0.0735$, $0.0817$, $0.0898$.	
	}
\label{fig:qx}
\end{figure}

The overlap $q(x)$ for different temperatures is
shown  in Figure (\ref{fig:qx}). The transition 
between $T=0.0817$ and $T=0.0898$ is easily recognizable
from the deviation of $q(x)$ from a constant
(the critical value  at which  the RS
solution breaks down is $T_c = 0.089725$). 
The rounding near the plateaus is an artifact 
of finite $N_x$. Indeed for increasing $N_x$ the shoulder becomes steeper
and steeper and, in the limit $N_x\to\infty$,
  $\dot q(x)$ develops a discontinuity
at the end points of the plateaus \cite{Tommy}.
By varying the extrema of the $x$-integration and $N_x$ the plateaus 
end-points $x_1$ and $x_2$ 
can be precisely identified. On then concludes that
the functional form of $q(x)$ is similar to the one
of the SK model in external magnetic field:
\begin{equation}
q(x)=\left\{\begin{array}{lr}
q(0) & 0 \leq x < x_1 \\
\mbox{non trivial} & x_1 < x < x_2
 \\
q(1) & x_2 < x \leq 1
\end{array}\right.
\end{equation} 
The analytic form of the non trivial part of $q(x)$
could be obtained from the resummation of high order expansions 
of the $\infty$-RSB equations 
similarly to what is done for the SK model \cite{Tommy}.
However, since observables such as
$q(0)$, $q(1)$, energy, etc., are not very sensitive (difference of
the order of the numerical precision) to the smoothness
of $q(x)$ we did not performed such an analysis here. 

\begin{figure}[hbt]
\begin{center}
\includegraphics*[width=0.7\textwidth,height=0.5\textwidth]{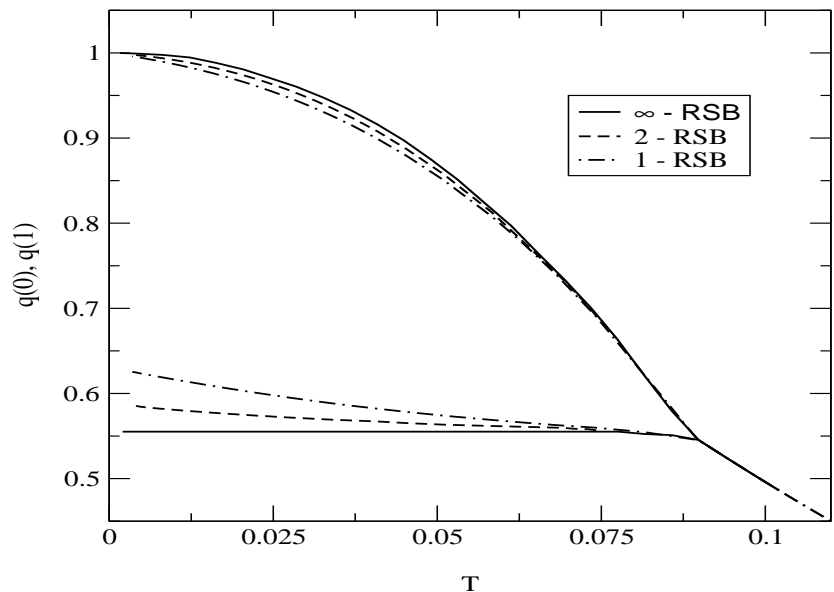}
\end{center}
\caption{The order parameter at the slowest ($x=0$)
	and at the fastest ($x=1$) time scales as a function of 
	reduced temperature.
	As the RSB scheme is improved the splitting between the two values 
	increases.
	}
\label{fig:q013}
\end{figure}

In Figure (\ref{fig:q013}) the behaviour of the largest
and smallest overlaps $q(1)$ and $q(0)$ as function of temperature 
is compared with the results from one and two RSB solutions.

For lower temperatures we used the Sommers' gauge which takes 
an anomaly with constant derivative \cite{SDJPC84}. 
At difference with the 
SK, here we have two anomalies and hence two possible choices.
However, the more natural one for numerical integration
$\dot\Delta_\lambda(x) = -\Delta_\lambda(0) = const$ 
leads to a more involved determination of $\Delta_\lambda(0)$.
Indeed we should first find $\dot\Delta_q(x)$ from eq. (\ref{Dq}) and then
$\Delta_\lambda(0)$ from (\ref{Dla}). Therefore for low temperatures
we adopted the Sommers' gauge $\dot\Delta_q(x) = -\Delta_q(0) = const$ ,
where [see eq. (\ref{Dq})]:
\begin{equation}
\Delta_q(0)= -\beta\bigl[1-q(1)\bigr]+
\int_{-\infty}^{\infty} dy\, P(0,y)\, m'(0,y)
\label{Dq0T}
\end{equation}
and $\dot\Delta_\lambda(x) = -\Delta_q(0)\, g_2(q(x))$.
We note that since $\dot\Delta_\lambda$ does not vanish for $x\to 1$
this leads to numerical instabilities for large temperatures.
Therefore from the point of view of numerical integration the two gauges 
are complementary. 

The order parameters $q(x)$  and
$\lambda(x)$ are different if we use the Parisi's
or the Sommers' gauge, but the thermodynamics observables are,
of course, invariant. This fact has been used to check the 
numerical integration by comparing the results from the two gauges
in the temperature range where both are stable.

\begin{figure}[hbt]
\begin{center}
\includegraphics*[width=0.7\textwidth,height=0.5\textwidth]{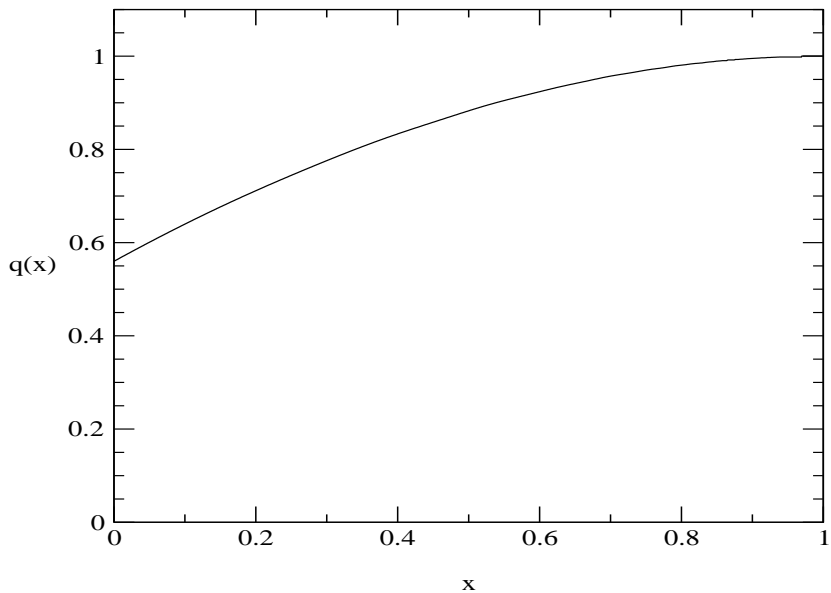}
\end{center}
\caption{The order parameter $q(x)$ at zero temperature for the $3$-SAT, 
	 in the gauge $\dot\Delta_q=-\Delta_q(0)$.
	}
\label{fig:qxT0}
\end{figure}

One of the main advantage of Sommers' gauge is that we can solve the 
equations at exactly $T=0$. In Figure (\ref{fig:qxT0}), for example, 
we report $q(x)$ for $T=0$ in the Sommers' gauge. We recall that 
in the Parisi's gauge $q(x) = q(1)$ for $x>0$ but $q(0)\not= q(1)$,
as can also be inferred from Figure (\ref{fig:qx})

For what concerns the thermodynamic quantities
one sees that 
using (\ref{q}) and (\ref{rel0}) the entropy must be 
proportional to $T^2$ for $T\to 0$ and hence vanishes. Moreover 
it also follows that in the same limit $q(1)\simeq 1-a T^2$
{\cite{SDJPC84,TAP}}.

It can be easily checked that
\begin{equation}
\lim_{T\to 0}\left[f_0+\frac{\beta}{2}g\left(q(1)\right)\right]=
\lim_{T\to 0}\left[
   \frac{\partial \beta f_0}{\partial\beta}+\beta g\left(q(1)\right)\right]=0
\end{equation}
so that the energy density for $T=0$ can be written as
\begin{equation}
 u =f_v= -\frac{1}{2}\int_0^1 dx \ q(x)\ \dot\Delta_{\lambda}(x)
 -\int_{-\infty}^{\infty} dy\ P(0,y)\ \phi(0,y)\ .
\label{fS1}
\end{equation}
The last term can be expressed as function of the local magnetization
using the identity
\begin{equation}
\phi(0,y)=\phi(0,0)+\int_0^y dy_1\ m(0,y_1)
\label{phi0y}
\end{equation}

where 
\begin{eqnarray}
\phi(0,0)&=&\int_0^{\infty} dy \left(1-m(0,y)\right)
-\frac{1}{2}\int_0^1 dx\ \dot\Delta_{\lambda}(x)
\nonumber
\\
&=&\int_0^{\infty} dy \left(1-m(0,y)\right)
+\frac{1}{2}\Delta_{q}(0)\int_0^1 dx g_2\left(q(x)\right) \ .
\label{phi00}
\end{eqnarray}

Using the relations derived in the previous section, alternative expressions
for $u$ can be obtained. For example by means of (\ref{rel1}) evaluated for
$T=0$ the energy density takes the form
\begin{equation}
f=u=-\int_0^1 dx \ q(x)\ \dot\Delta_{\lambda}(x)
-2 \int_{0}^{\infty} dy\ y\ P(1,y) \ .
\label{fS2}
\end{equation}
This can be simplified further using relation (\ref{rel2}), which in
the chosen gauge at $T=0$ becomes 
\begin{equation}
\Delta_q(0)
\int_0^1 dx\ \left[q(x)\ g_2\left(q(x)\right)+g_1\left(q(x)\right)\right]
=2 \int_0^{\infty} dy\ P(1,y)\ y\ .
\label{Dq0}
\end{equation}
so that (\ref{fS2}) takes the form:
\begin{equation}
\fl u=\Delta_q(0)\int_0^1 dx \ q(x)\ g_2\left(q(x)\right)
- 2 \int_{0}^{\infty} dy\ y\ P(1,y)=
-\Delta_q(0)\int_0^1 dx\ g_1\left(q(x)\right) \ .
\label{u0S}
\end{equation}
For the 3-SAT problem this reads:
\begin{equation}
u=-\frac{1}{128}\Delta_q(0)\int_0^1 dx\ \left(1+q(x)\right)^2 \ .
\label{u0S3}
\end{equation}

\begin{figure}[hbt]
\begin{center}
\includegraphics*[width=0.7\textwidth,height=0.5\textwidth]{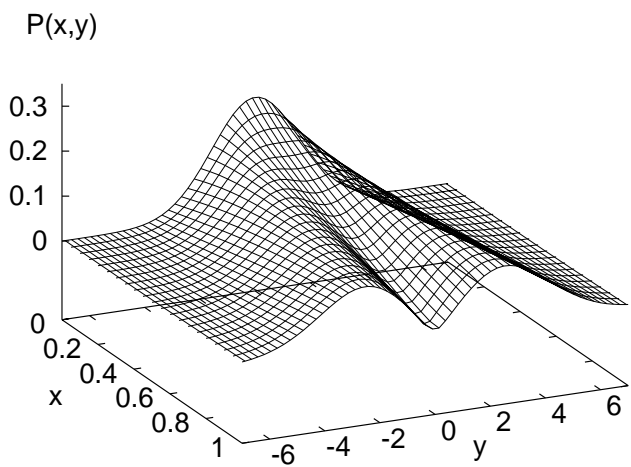}
\end{center}
\caption{The local field probability distribution $P(x,y)$ for
	 the $3$-SAT at T=0 in the Sommers' gauge.
        }
\label{fig:PhyST03}
\end{figure}

We conclude this section showing in Figure (\ref{fig:PhyST03})
the probability distribution 
$P(x,y)$ of frozen fields at $T=0$ for different time scales $\tau_x$.
From the figure it is evident that the distribution of the field $y$
varies continuously from a Gaussian, for very short time-scales, 
to a double peak distribution, for the longest time-scales.

\section{Thermodynamics of  the highly constrained 3-SAT problem}
\label{TD}

\begin{figure}[hbt]
\begin{center}
\includegraphics*[width=0.7\textwidth,height=0.5\textwidth]{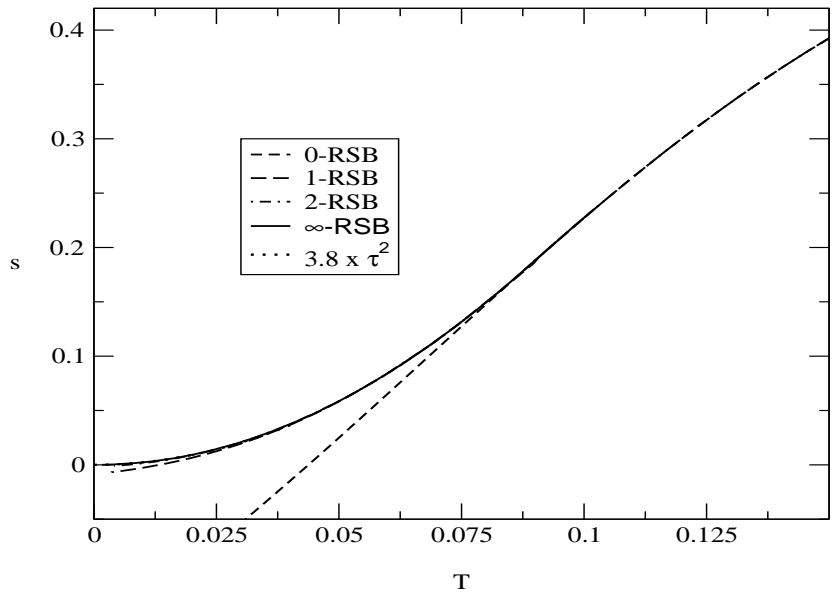}
\end{center}
\caption{Entropy density of the $3$-SAT model for large number of clauses. 
	The Replica Symmetric, $1$-RSB, $2$-RSB and $\infty$-RSB solutions 
	are plotted. For the latter $s(0)=0$. On this scale the $2$-RSB
        and $\infty$-RSB are almost indistinguishable.
        }
\label{fig:entro3}
\end{figure}

In Figure (\ref{fig:entro3}) we show the entropy density 
as a function of  temperature down to $T=0$. For each temperature, 
including $T=0$, the data are obtained using the  gauge appropriate
for that temperature.
For comparison, the entropy computed within the 
Replica Symmetric, $1$-RSB and $2$-RSB solutions \cite{LPJSP01}
are also plotted.
As it can be seen from the log-lin plot, the $2$-RSB solution is a very good 
approximation but yet it is inexact when $T<0.016$.
The entropy is zero for $T=0$ as confirming the conjecture of Ref.
\cite{MZPRE97} for the behaviour in the UNSAT-phase.
As expected,  $s$ vanishes quadratically with the temperature.

\begin{figure}[hbt]
\begin{center}
\includegraphics*[width=0.7\textwidth,height=0.5\textwidth]{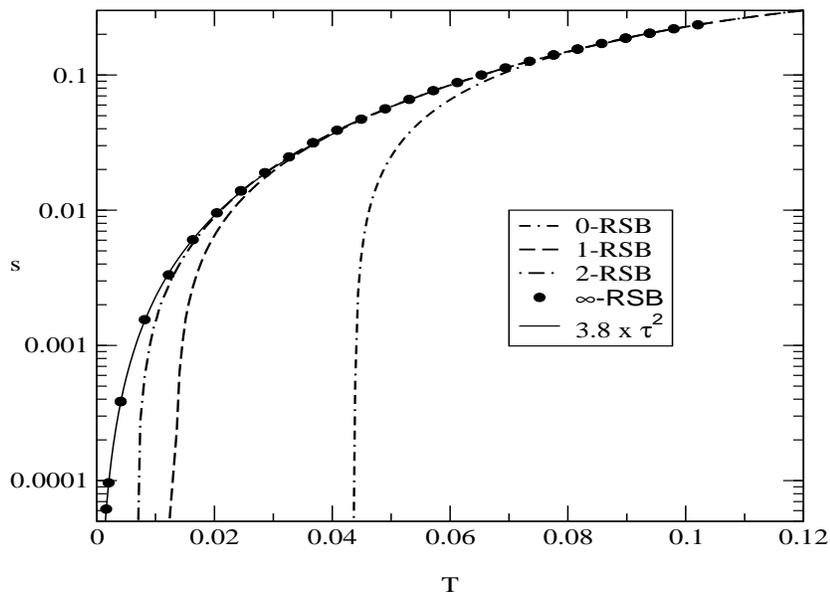}
\end{center}
\caption{Same as Fig. (\protect{\ref{fig:entro3}})
	 The improvement got in the low temperature
         region by breaking 
	 the replica symmetry is clearly seen.
	 The data for $\infty$-RSB are reported as circle to distinguish
	 them from the quadratic fit.
        }
\label{fig:entro3l}
\end{figure}

\begin{figure}[hbt]
\begin{center}
\includegraphics*[width=0.7\textwidth,height=0.5\textwidth]{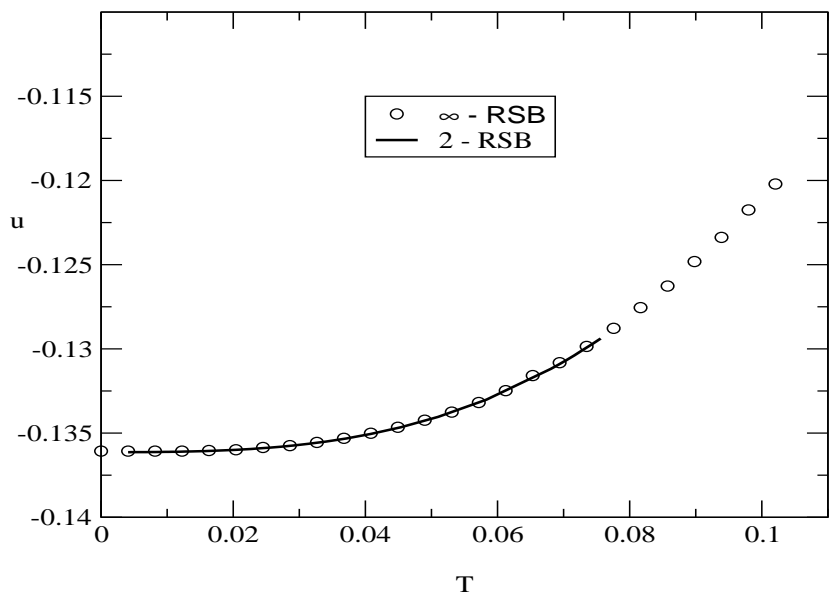}
\end{center}
\caption{Energy density for the 3-SAT problem at high 
	 connectivity ($\alpha\gg\alpha_c$) in the 
	 2-RSB and $\infty$-RSB solutions.
	 The coincidence between the 2-RSB energy and the 
	 $\infty$-RSB energy is valid up to order $10^{-4}$.
        }
\label{fig:ene3}
\end{figure}

Finally in Figure (\ref{fig:ene3}) we show the energy density.

The quantity plotted is actually 
$u(T)-\sqrt{\alpha}/48$, where $\alpha$ is very large.
Equating the internal energy to zero we can determine an upper
bound for the critical value $\alpha_c$ of the ratio
of the number of clauses to the number of variables that marks the transition
between the UNSAT-phase (in which we derived our asymptotic model) and the
SAT-phase, where the energy is, by definition, always zero, at $T=0$.
Using the $\infty$-RSB solution we get $\alpha_c^{\rm u.b.}=7.10969$.
For the $2$-RSB solution it was already 
$\alpha_c^{\rm u.b.}=7.11400$ \cite{LPJSP01}.

\section{Conclusions}
We performed the study of the Replica Symmetry Breaking solutions
of the 3-SAT problem in the limit of many clauses, mapping it in a
poorly diluted spin glass model with long-range random quenched interactions.
The mapping to a statistical mechanics model was carried out
 introducing an artificial temperature and taking, in the end, the limit
$T\to 0$, to recover the original model.
We found that the structure of the solutions to the problem
is of the $\infty$-RSB kind: in order to get a stable solution
the replica symmetry has to be broken in a continuous way, similarly 
to the SK model \cite{SKPRL75} (in external magnetic field).
The $\infty$-RSB structure holds down to the interesting limit
of zero temperature.

No phase transition is expected in the UNSAT phase, other than 
the SAT-UNSAT transition occurring at $\alpha=\alpha_c\simeq 4.2$ 
Therefore we expect the same $\infty$-RSB structure of solutions of
found for the over-constrained case to hold
also in the critical region.

From the value of the energy at zero temperature we find 
the upper bound $\alpha_c < 7.10969$. to the critical value of the 
number of clauses per variable.
Even if this is of the same order of magnitude of
$\alpha_c\simeq 4.2$ \cite{SK96} yielded by direct numerical simulations, 
it is still too large.
We recall that such a value has been obtained
through a first order expansion 
in $1/\sqrt{\alpha}$. 
In order to get a better approximant other terms
should be considered, possibly more then one since we are dealing with 
an asymptotic expansion and therefore nothing guarantees that 
the second order corrections are small and in the right direction.

Finally,
as by product, in the present paper we worked out a
precise  procedure to get the $\infty$-RSB solution of a 
general class of models that, besides the over-constrained 3-SAT model, 
include SK,
p-spin and, more generally, models with  
any combination of $p$ interacting terms.
We presented the solution exploiting a variational method, introduced by 
Sommers and Dupont \cite{SDJPC84}, which has the advantage of
being easily implemented on a computer for any temperature
including $T=0$.
As a consequence the numerical code developed to solve the present 
model can be applied to the whole
class of models without any relevant change, providing an efficient tool
for the analysis of the structure of the solutions of a large number of
spin models interacting via quenched random couplings.

\label{concl}

\end{document}